%% file: main.tex
\documentclass{article}
\usepackage[utf8]{inputenc}
\usepackage{eurosym}
\usepackage{amstext} 
\DeclareRobustCommand{\officialeuro}{%
  \ifmmode\expandafter\text\fi
  {\fontencoding{U}\fontfamily{eurosym}\selectfont e}}
\usepackage{nomencl}
\usepackage{etoolbox}
\usepackage{amsmath}
\usepackage{amssymb}
\usepackage{array}
\usepackage{mdwmath}
\usepackage{mdwtab}
\usepackage{array}
\usepackage{graphicx}
\usepackage{appendix}
\usepackage{cleveref}
\usepackage{textcomp}
\usepackage{caption}
\usepackage{subcaption}

\usepackage[left=3cm, right=3cm, top=2.5cm]{geometry}

\renewcommand\nomgroup[1]{%
	\ifstrequal{#1}{I}{\item[\textbf{Indices and Sets}]}{%
	\ifstrequal{#1}{P}{\vspace{10pt} \item[\textbf{Parameters}]}{%
	\ifstrequal{#1}{V}{\vspace{10pt} \item[\textbf{Variables}]}{}}}%
	\ifstrequal{#1}{W}{\vspace{10pt} \item[\textbf{Dual variables}]}{}
}

\title{Grid Tariffs Based on Capacity Subscription: Multi Year Analysis on Metered Consumer Data}
\author{Sigurd Bjarghov$^{a*}$, Hossein Farahmand$^a$, Gerard Doorman$^{b}$ \\
$^a$Department of Electric Power Engineering, NTNU -- Trondheim, Norway\\
$^b$Statnett -- Oslo, Norway\\
$^{*}$ Corresponding author - Email:\ sigurd.bjarghov@ntnu.no}

\date{}

\begin{document}

\maketitle

\makenomenclature
\setlength{\nomlabelwidth}{3cm}
\nomenclature[I]{$T$}{Set of time steps, index $t$}
\nomenclature[I]{$S$}{Set of load scenarios, index $s$}
\nomenclature[I]{$J$}{Set of value of cut load segments, index $j$}

\nomenclature[P]{$L_{ts}$}{Load [kWh/h]}
\nomenclature[P]{$C^{fix}$}{Annual fixed grid tariff cost [$\frac{ \euro }{year}$]}
\nomenclature[P]{$C_j^{VCL}$}{Value of cut load [$\frac{ \euro }{kWh}$]}
\nomenclature[P]{$C^{l}$}{Grid tariff energy cost [$\frac{ \euro }{kWh}$]}
\nomenclature[P]{$C^{h}$}{Grid tariff excess energy cost [$\frac{ \euro }{kWh}$]}
\nomenclature[P]{$C^{sub}$}{Grid tariff subscribed capacity cost per kW per year [$\frac{\euro }{kW-year}$]}

\nomenclature[V]{$x_{ts}^{l}$}{Bought electricity below sub. cap. [kWh/h]}
\nomenclature[V]{$x_{ts}^{h}$}{Bought electricity above sub. cap. [kWh/h]}
\nomenclature[V]{$x^{sub}$}{Subscribed capacity [kW]}
\nomenclature[V]{$x_{tsj}^{VCL}$}{Cut load in segment $j$ [kWh/h]}

\printnomenclature

\begin{abstract}

While volume-based grid tariffs have been the norm for residential consumers, capacity-based tariffs will become more relevant with the increasing electrification of society. A further development is capacity subscription, where consumers are financially penalised for exceeding their subscribed capacity, or alternatively their demand is limited to the subscribed level. The penalty or limitation can either be static (always active) or dynamic, meaning that it is only activated when there are active grid constraints. We investigate the cost impact for static and dynamic capacity subscription tariffs, for 84 consumers based on six years of historical load data. We use several approaches for finding the optimal subscription level ex ante. The results show that annual costs remain both stable and similar for most consumers, with a few exceptions for those that have high peak demand. In the case of a physical limitation, it is important to use a stochastic approach for the optimal subscription level to avoid excessive demand limitations. Facing increased peak loads due to electrification, regulators should consider a move to capacity-based tariffs in order to reduce cross-subsidisation between consumers and increase cost reflectivity without impacting the DSO cost recovery.

\end{abstract}

\input{Chapters/Introduction.tex}
\input{Chapters/Capacitytariffs.tex}

\input{Chapters/Model.tex}

\input{Chapters/Casestudy.tex}

\input{Chapters/Results.tex}
\input{Chapters/Conclusion.tex}
\input{Chapters/Acknowledgements.tex}
\input{Chapters/Appendix.tex}

\bibliographystyle{IEEEtran}
\bibliography{IEEEabrv,mylib,references}
\end{document}

%% file: Chapters/Introduction.tex
\section{Introduction} \label{introduction}

\subsection{Background}

As a measure to reach emission targets, Norway is considering a significant increase in electricity consumption by electrifying transport, off-shore installations and various industries \cite{NVE2020, Statnett2020}. Meanwhile, household peak loads are expected to increase due to charging of electric vehicles (EVs) and electrification of heating. This might increase peaks loads in parts of the grid, which could result in significant expected grid investments in coming years. 

Against the backdrop of these developments and with the intention to reduce grid investments, the Norwegian regulator (RME) proposed several new grid tariff structures to incentivise demand response during peak load hours. One of the suggestions by the Norwegian regulator, is a capacity subscription (CS) tariff, where customers subscribe to a capacity level. Similar to current grid tariff structures, it contains an annual fixed cost reflecting the distribution system operator's (DSO) fixed costs. In addition, there is a capacity cost (per kilowatt), with some resemblance to an internet subscription with a specific bandwidth speed. Demand below the subscription level has a small energy term, which reflects marginal grid losses, whereas demand above has a high energy term, which penalises excess use. This CS tariff thus incentivises customers to keep their demand below the subscribed capacity level. The tariff is "static" in the sense that it always penalises excess consumption, regardless of whether the grid has any congestions. This is sub-optimal in terms of cost-reflectiveness, as consumers are penalised for their peak demand whether there is a system peak or not.

To address the issue of system versus consumer peak coincidence, we suggest a ``dynamic'' CS as an alternative, where capacity limits are only activated when there is scarcity of grid capacity. In the case of scarcity, consumers are physically limited to their subscribed capacity using load limiting devices (LLD), unlike the static version where only an excess energy term has to be paid\footnote{It would also be possible to use a ``financial'' version, which would include payment for excess demand like in the static tariff. The excess cost coefficient would be higher because activation is done only sporadically.}. This dynamic capacity subscription concept is more efficient because there is no penalty for using capacity when there is no scarcity. This form for capacity subscription was first presented in \cite{Doorman2005}, but there it focused on the power market instead of the grid. ``Activations'' of the LLDs would be done by the DSOs, which issue warnings in advance. Customers then have a reasonable amount of time to plan for reducing demand during the activations, by demand response or distributed energy resources (DER). Smart meters make it possible not only to analyse demand patterns to find a fitting subscription limit but also to implement such a grid tariff structure. 

In previous work by the authors, ``static'' and ``dynamic'' CS tariffs were analysed for a single customer for one year \cite{Bjarghov2018}, where the results showed mainly two things: sub-optimal static subscribed capacity levels do not critically influence annual costs, and the dynamic optimal subscribed capacity level depends heavily on amount of activations, which is unknown \textit{ex ante}. Furthermore, the need for a more extensive study covering more customers and uncertainty in demand was highlighted. Capacity subscription was also found to be the market design that was closest to optimum and leads to highest surplus on the consumer side rather than the supply side \cite{Doorman2008AnalysisDesigns}. 

The issue of social fairness has been raised in response to the suggested network tariff change. Although new price signals and incentives to shift towards more grid-friendly demand profiles can result in reduced socioeconomic costs, this may have undesirable distributional side effects. In essence, customers with grid-friendly demand profiles should have reduced costs and vice versa, given a properly designed, cost reflective grid tariff structure. The EU commission has highlighted in the ``Clean Energy For All Europeans'' package \cite{EUcommission} that \textit{"The package also contains a number of measures aimed at protecting the most vulnerable consumers”}. It is thus important to consider if vulnerable consumers could be harmed by the proposed tariff change. Further, economic efficiency is not the only criterion of a grid tariff design. For the DSO, cost-recovery and stability of annual revenues are particularly important. On the consumer side, fairness and acceptance are considered to be of high importance. These qualitative criteria are challenging to define, which makes grid tariff design a difficult task \cite{Brown2015EfficientServices, PerezArriaga2016UtilityResponse}. Still, the change from energy-based to capacity-based tariffs is also taking place in the Dutch speaking part of Belgium, Flanders, which will introduce a capacity-based grid tariff from mid 2022. For households and small companies, this is based in the rolling-average monthly 15-minute peak, with a minimum value of 2.5 kW \cite{VREG}.

\subsection{Literature overview}

Fairness-related issues of grid tariffs have also been discussed in recent literature. A redistribution of costs between residential consumers was shown in \cite{Saele2017}, where up to 15\% of the costs were shifted from consumers with low peak loads to consumers with high peak loads. \cite{Hledik2016TheCharges} finds that demand charges do not disproportionately impact low-income customers relative to rest of sample, the results show that a demand charge tariff will change the bill of roughly 10 \% of the customers by more than 20\%. Further, \cite{Burger2020} points out that a two-part tariff mitigates the potential average increase in tariff costs for low-income customers. Capacity tariffs are found to be more fair than flat, peak and Ramsey pricing in \cite{Neuteleers2017AssessingTariffs}. Although working well, capacity-based tariffs might lead to over-investments in demand response or other types of flexibility which might lead to other competition-related issues where flexibility owners push costs over on other customers \cite{Schittekatte2018Future-proofBack}. 

An advantage of capacity-based tariff structures is the removal of cross-subsidisation of distributed generation, which is an increasing issue with the rapid increase in photovoltaic panels \cite{Schreiber2013Capacity-dependentSystems, Hledik2014RediscoveringCharges, Picciariello2015ElectricityProsumers}. \cite{Jargstorf2013Capacity-basedCustomers} also claimed that tariffs were more efficient with a fixed, energy-based and capacity-based share to reduce cross-subsidies. 

Demand charges are relatively common for commercial and industrial customers. With the use of smart metering, the peak load of a certain time period (typically monthly) is measured and the consumer pays per kilowatt or megawatt peak. The authors in \cite{Schreiber2013Capacity-dependentSystems}, observe that demand charges (like in Flanders \cite{VREG}) have the "early peak" issue where an early peak in a monthly measured network tariff structure removes incentives for reducing peak loads for the rest of the tariff period. This does not occur with CS tariffs as the excess energy term applies for all peaks above the subscription level. Further, \cite{Bartusch2011IntroducingPerception} showed that consumers were relatively positive to demand-based tariffs under the assumption that it is easy to monitor your consumption.

Like the static version of the capacity subscription tariff, demand charges are inefficient if the customer peak does not coincide with system peak \cite{Borenstein2016TheUtilities}. There has also been claims that energy-based tariff costs correlate strongly with peak-demand, suggesting that demand-based tariffs are unnecessary \cite{Blank2014ResidentialDebate}. This is supported by \cite{Borenstein2016TheUtilities} in which the authors also question whether demand charges are cost-reflective as system and consumer peak do not necessarily coincide. However, these claims were made before the increase in residential peak loads seen in countries with a high share of EVs \cite{Saele2018}.

Developments towards lower marginal costs of energy and more capital intensive technologies presently increase interest in solutions based on capacity subscription, both for energy and grid tariffs. Lack of capacity in distribution grids was highlighted as an important barrier for electric vehicle (EV) integration in Norway. In addition to the authors' previous work, \cite{Backe2020ComparingTariffs,Pinel2019,Almenning2019} pinpoint that the coincidence factor of consumer versus system peaks can be dealt with by forming energy communities (under a neighbourhood tariff). Similar results are achieved in \cite{Hennig2020CapacityManagement}, which showed an increased capability of integrating EVs into the distribution grid under a CS tariff scheme. Also under competitive, local electricity market schemes, the market is able to flatten peak loads \cite{Bjarghov2020}. An example is shown in \cite{Askeland2021ActivatingNorway}, where the concept of EV integration was demonstrated in a real case study in Norway, where a neighbourhood were able to adopt more EVs by coordinating flexible resources under capacity-based tariffs. Consumers adapting to grid tariffs is an apparent consequence of higher distributed energy resources shares in the future.  

\subsection{Contributions \& paper organisation}

The purpose of this paper is to investigate the economic impact and efficiency of CS tariffs on consumers and DSO. Therefore, we analyse the impact on passive consumers (with no production nor flexibility) under static and dynamic CS tariffs for 84 customers with six years of demand data. The main contributions of this paper are the following:
\begin{itemize}
    \item We analyse the economic impact of static and dynamic capacity subscription grid tariffs for larger sample of consumers over multiple years.
    \item We propose a method to determine the optimal subscription level based on a stochastic approach and demonstrate the advantages of this method compared with the naive approach of using the previous year's data.
    \item We demonstrate how many consumers that experience significant cost deviations from capacity subscription tariffs compared with existing tariffs, in relation to their relative peak loads.
    \item Under dynamic capacity subscription, where demand is limited only when there is a grid scarcity, we investigate the difference in how much capacity consumers procure to avoid excessive demand limitations compared to the static variant, modelled by an assumed discomfort function.
\end{itemize}

The remainder of the paper is organised as follows: \Cref{capacitytariffs} discusses the CS grid tariff design. The model is presented in \Cref{model}, followed by the case study description in \Cref{casestudy}. Results and discussions are then presented in \Cref{results}, followed by conclusions and further work suggestions in \Cref{conclusion}.

%% file: Chapters/Capacitytariffs.tex
\section{Capacity tariffs}\label{capacitytariffs}

\subsection{Static capacity tariff}

The capacity subscription tariff proposed by the regulator has four components: a fixed annual cost (€), a capacity cost (€/kW), an energy cost (€/kWh) and an excess demand charge (€/kWh). Note that, in addition to the grid tariff, the consumer pays for electricity and taxes, but in this paper we only focus on the grid tariff. The annual consumer grid cost is calculated as shown in \eqref{annualcustomercost}.

\begin{equation}
C^{tot} = C^{fix} + C^{sub} \cdot x^{sub} + \sum_t (C^l \cdot x_t^l + C^h \cdot x_t^h)
\label{annualcustomercost}
\end{equation}

Where $C$s are the various cost coefficients explained above, $x^{sub}$ is the subscribed capacity, $x^l$ the annual consumption below the subscribed capacity level and $x_{h}$ the demand in excess of the subscribed capacity. $C^l$ should cover the average losses in the grid and is typically around 0.5€ct/kWh. Because $C^h$ is significantly higher, the consumer has an incentive to keep demand below the subscribed capacity, $x_{sub}$. 

According to the regulation, the grid companies that apply CS tariffs will be obliged to recommend the $x^{sub}$ minimising $C^{tot}$ to the consumer. Because hourly demand data will be available, this is in principle an easy task based on ex post data. An updated proposal required that the last 12 months are used for determining the subscription level; it is changed dynamically each month. In our analyses, we will find the optimal $x^{sub}$ based on six years of data, but we will also look at other ways to find $x^{sub}$. 

In this study, customers can subscribe to any capacity, whereas, in reality, it is reasonable to assume that customers have to choose between discrete steps with e.g. 0.25, 0.5 or 1 kW intervals. A high resolution of choices makes it more complicated for customers to choose, whereas a low resolution, with e.g. 1 kW intervals would create sub-optimal conditions for customers with a low consumption due to a high deviation between optimal subscribed capacity (e.g. 1.5 kW, and the choices that would be 1 or 2 kW). This is less relevant as annual demand (and thus average demand) increases. We abstract from this issue and assume a continuous scale in our study to get a more precise idea of which subscription levels are optimal. 

One of the design parameters of subscription-based tariffs is the frequency of subscription level updates. From the perspective of the DSO, annual subscription might be preferable, especially when demand is strongly influenced by seasonal variations. On the other hand, consumers need flexibility with respect to changing circumstances. Examples of changing circumstances that heavily influences the optimal subscription limit could be moving or investments in demand increasing/decreasing assets such as EVs, house insulation upgrades or heat pumps. 

In this paper, we (among other approaches) investigate a subscription level which is decided annually ex ante. However, in a real implementation it must be possible to adjust the level during the year, without allowing consumers to subscribe to a low level in a typical low load season (summer in a cold climate) and then increase subscription during a high load season. If this were allowed, capacity prices would need to be adjusted correspondingly. The approach proposed by the Norwegian regulator, to base subscription on demand during the last 12 months, updated on a monthly basis, solves the problem of the frequency of update, but is sub-optimal as we will show in this paper. Moreover, it only partly takes into account major changes in demand, which will only slowly result in a corresponding change in subscribed capacity. Another possibility is that capacity is paid for on an annual basis, but that is a secondary market for shorter commitment periods. We do not elaborate on this issue in this paper, but it is an open issue for further research.

\subsection{Dynamic capacity tariff} \label{dynamiccapacitytariff}

Capacity subscription was proposed in \cite{Doorman2008AnalysisDesigns} for the power market. In \cite{Doorman2017ElectricityCapacity}, the authors also indicated the possibility to use the same model for the grid tariff structure. An essential feature of the dynamic CS is that demand is limited to the subscribed capacity when there is scarcity in the system (i.e. not enough generation capacity in the ``market case'' or an active grid constraint in the present context). In such cases, the DSO (or TSO) activates a Load Limiting Device (LLD), effectively limiting demand. To make this acceptable for the consumer, it is necessary to have intelligent load control that keeps demand below the subscribed limit, by switching off non-essential demand like floor heating or other appliances. Delaying the charging of EVs is also very well suited to keeping demand right below the limit. Here we use the term ``dynamic'' CS, to distinguish it from the tariff proposed by the Norwegian regulator \cite{NVE2017}. The consumer cost is very similar to equation \eqref{annualcustomercost}, but there is no excess consumption above the subscription level, because demand is limited instead. On the other hand, the consumer experiences a partial loss of load, which in effect is a welfare loss that needs to be considered in the cost optimisation. The annual customer total cost under the dynamic CS tariff scheme is presented in \eqref{dynamicannualcustomercost}.

\begin{equation}
C^{tot} = C^{fix} + C^{sub} \cdot x^{sub} + \sum_t (C^l \cdot x_t^l + \sum_j C_{j}^{VCL} \cdot x_{tj}^{VCL})
\label{dynamicannualcustomercost}
\end{equation}

The costs are very similar to the static CS tariff, but instead of an excess energy term, consumers experience a discomfort cost ($C_{j}^{VCL}$ which increases the more load $x_{tj}^{VCL}$ is cut. Because the discomfort costs increases exponentially as more load is cut, this is segmented (indexed by $j$) in a piecewise linearised fashion. The discomfort costs are discussed in detail in \Cref{subsec:discomfortcosts}.

\subsubsection{Discomfort costs} \label{subsec:discomfortcosts}

To determine the optimal subscription level for dynamic CS, the consumer cost of having to reduce load must be taken into account. This cost cannot be observed, like the excess demand cost $C^h$ under the static CS tariff. The value of lost load typically depends on customer type and duration of disconnection, and represents the discomfort costs of electricity not served. However, under the dynamic CS scheme, the load is only limited and not completely disconnected. As stated in \Cref{dynamiccapacitytariff}, intelligent load control can be utilised, disconnecting only non-essential demand, which further leads to lower comfort loss. The value of cut load ($C^{VCL}$) is a function of how much load is disconnected. We use the formulation in \eqref{VCLequation}, which was also used in \cite{Bjarghov2018}.

\begin{equation}
C^{VCL} = \frac{VoLL}{1-e^{-b\cdot L}}(1-e^{-b(L-x^{sub})})
\label{VCLequation}
\end{equation}

\begin{figure}[h]
    \centering
    \includegraphics[width=0.9\columnwidth]{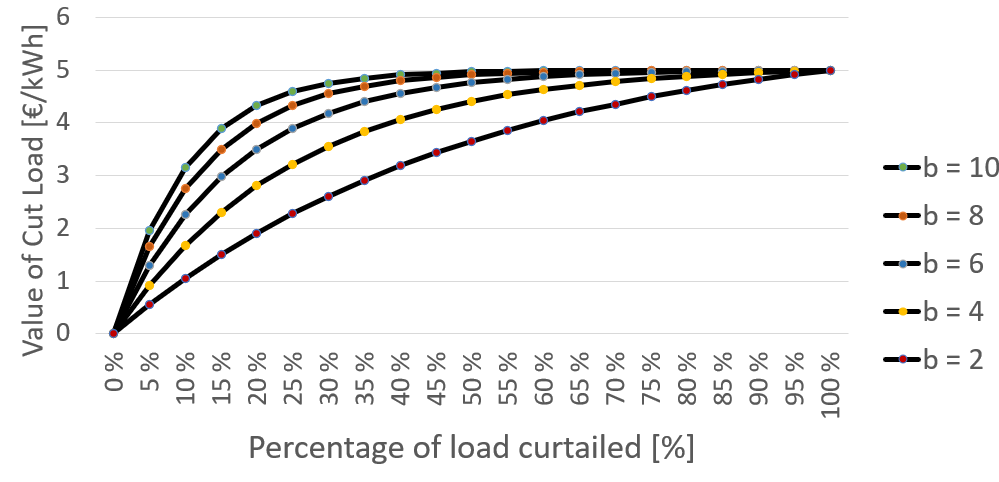}
    \caption{Value of cut load with different steepness levels. VoLL is 5 \EUR{}/kWh.}
    \label{VCLpercent}
\end{figure}

Value of cut load is represented as a value between 0 and VoLL as a non-linear curve as demonstrated in \eqref{VCLequation}. The curve steepness is given by $b$. The load $L$ and subscribed capacity level $x^{sub}$ decides the discomfort cost in a certain time period. If $L$ never exceeds $x^{sub}$ (which translates to subscribing to the maximum demand), discomfort costs will be zero. The impact of different values of $b$ is visualised in \Cref{VCLpercent}. A steep VCL curve (high $b$) translates to the consumer having high discomfort costs of curtailing a relatively low share of the consumer's load. A low steepness (low $b$) implies that the consumer is quite flexible and can curtail more load without experiencing high discomfort. In this paper, we assume a steepness $b$ of 8, resulting in a relatively steep discomfort cost curve, as shown in \Cref{VCLpercent}. This level implies that the cost of flexibility (and thus discomfort costs) is relatively high, and also only results in small reductions in load. In reality, this level could be adapted to each individual customer based on their real discomfort costs.

If a consumer has a peak load of 5 kW, the curtailment of 1 kW (20\%) should have a similar discomfort cost as a consumer with a 10 kW peak load who is curtailed 2 kW (also 20\%). This is a necessary simplification made to be able to compare curtailment of different customer types. In this approach, $C_j^{VCL}$ is decided based on the maximum load of the consumer in the specific year that is simulated. The consequence is that 1 kW of curtailment is not given the same discomfort cost for all consumers. The value of lost load depends on customer type, duration and time \cite{Schroder2015}, but we simplify by setting it to 5 \EUR{}/kWh. 

\subsubsection{Activation of capacity subscription}

An important advantage of the dynamics CS tariff is that it does not punish load above the consumer's subscription limit in non-scarce hours, and thus avoids the welfare loss caused by excess payment for a non-scarce resource. It also rewards flexible users who can reduce their consumption when the system requires it the most, or customers who simply do not have a high consumption when there is grid scarcity, as those consumers could subscribe to lower capacities and thus reduce their annul grid tariff costs. The system peak load varies from year to year because some winters are colder than others. The DSO will therefore invest in grid capacity that covers the highest peak in not only a year, but for several years. If there is a considerable penetration of electrical heating, this means that a winter without very cold periods could have no capacity scarcity, whereas years with cold periods would result in many hours with capacity scarcity. This makes it challenging to find a correct number of activations. If the DSO sets the threshold for activation relatively low, there could be so many activations that customers are incentivised to subscribe to a capacity close to their peak load to avoid high discomfort costs. The mentioned scenario would not increase social welfare much as the situation would remain quite similar as it is today where end-users indirectly pay to use whatever peak they want. If the threshold is high, the entire basis for choosing a subscription limit is based on very few hours per decade, which incentivises speculation in subscribing to 0 kW as well as demand response investments. This could make it difficult for the DSO to recover their existing costs, although they are reduced somewhat due a reduced peak load. To keep a good compromise between the two, there needs to be some activations each year, even if there is no severe capacity scarcity in the system.

Although the CS tariff primarily aims to reduce costly future distribution grid investments by incentivising customers to reduce peak loads, the DSO is not the only stakeholder here. The TSO who owns the transmission grid also has an interest in reducing future grid investments by avoiding growth in capacity use in areas with an increase in load. TSO and DSO interests therefore align because DSOs pay transmission grid rent in the hour where the total peak demand is the highest in the region. It is therefore realistic that both distribution and transmission grid-related congestions could lead to activations. Note that in this study we only base activations on local congestions.

Another important aspect of activations based on local congestions is if activations should be across the whole DSO grid (to avoid discrimination) or if they can be limited to overloaded radials only. Clearly, the latter would be the efficient solution, but it may contradict with rules on equal treatment. The present rules in Norway do not allow, e.g., different tariffs within the same DSO area, but it is not evident that a different number of LLD activations would fall under this requirement. However, this looks probable. On the other hand, it is clearly inefficient to activate LLDs across the whole DSO area because one or two radials are overloading, and this problem increases as the DSOs merge and become larger. A possible solution could be to reduce the fixed part of the tariff $C^{fix}$ for consumers on ``weak'' radials that can expect more LLD activations.

%% file: Chapters/Model.tex

\section{Model}\label{model}

\subsection{Stochastic optimal static subscribed capacity problem}

In Norway, the demand is weather dependent due to high penetration of electrical heating. Future demand is therefore unknown and varies from year to year. Therefore, choosing the optimal subscription level is a stochastic problem. We model this by using a number of historical weather years, represented by the index $s$, each having a probability $p_s$. Ideally, these years should present a statistically representative sample for the expected weather conditions. The consumer's objective function is given by \eqref{objfunsubcap}.

\begin{equation}
    \min{C^{fix} + x^{sub}C^{sub} + \sum_{s}\sum_{t}p_s (x_{ts}^{l}C^{l} + x_{ts}^{h} \cdot C^{h}}) \label{objfunsubcap}
\end{equation}

Energy bought from the grid is split into energy below ($x_{ts}^{l}$) and above ($x_{ts}^{h}$) the subscribed capacity $x^{sub}$ in \eqref{subcap1} and \eqref{subcap2}.

\begin{equation}
     x_{ts}^{l} + x_{ts}^{h} = L_{ts} \quad \forall t,s
     \label{subcap1}
\end{equation}

\begin{equation}
     x_{ts}^{l} \leq x^{sub} \quad \forall t,s
     \label{subcap2}
\end{equation}

\subsection{Stochastic optimal dynamic subscribed capacity problem} \label{stocdynsubcapsection}
In this case, the consumer's objective function is given by \eqref{objfundynsubcap}. The objective is straightforward, with an annual fixed cost, a subscription cost and an energy fee. In contrast to the static CS tariff, excess energy use is no longer possible, and the discomfort cost $C_j^{VCL}$ replaces $C^h$. Because consumers have different load profiles and annual demand, $C_j^{VCL}$ must be tailored for each consumer. The values used in these simulations are based on the curve presented in \Cref{VCLpercent}.

\begin{equation}
    \min{C^{fix} + x^{sub}C^{sub} + \sum_{s}\sum_{t} p_s {[(x_{t}^{l} \cdot C^{l} + x_{tsj}^{VCL} \cdot C_j^{VCL}]}} \label{objfundynsubcap}
\end{equation}

Energy from the grid is split into energy below ($x_{ts}^{l}$) the subscribed capacity $x^{sub}$ and the partially curtailed load $x_{tsj}^{VCL}$ above this capacity in \eqref{dynsubcap1} and \eqref{dynsubcap2}. 

\begin{equation}
     x_{ts}^{l} + \sum_j x_{tsj}^{VCL} = L_{ts} \quad \forall t,s
     \label{dynsubcap1}
\end{equation}

\begin{equation}
     x_{ts}^{l} \leq x_{t}^{sub} \quad \forall t,s
     \label{dynsubcap2}
\end{equation}

%% file: Chapters/Casestudy.tex
\section{Case Study} \label{casestudy}

\begin{figure}[h]
    \centering
    \includegraphics[width=0.9\columnwidth]{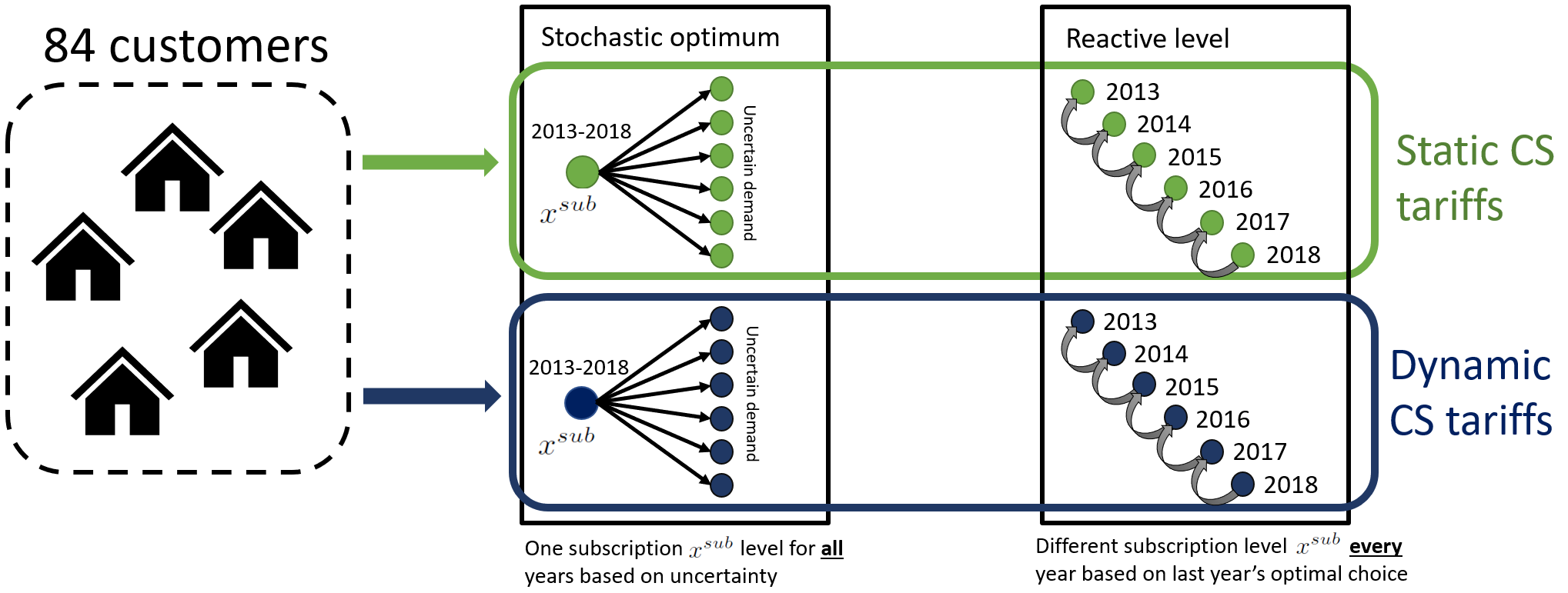}
    \caption{Overview of the case studies.}
    \label{caseoverview}
\end{figure}

Hourly load data from 84 customers in the NO1 price zone for the period 2013-2018 were used in the analysis. Because of privacy rules, the data could not be coupled to heating source or other information that might explain the load profiles. To analyse the impact of CS tariffs, the consumer costs under historical load data has been simulated with the existing energy tariff alongside the static and dynamic CS tariff schemes. An overview of all the simulations performed is presented in \Cref{caseoverview}. They are further explained in the subsequent sections.

\subsection{Customer data}

\begin{figure}[h]
    \centering
    \includegraphics[width=0.9\columnwidth]{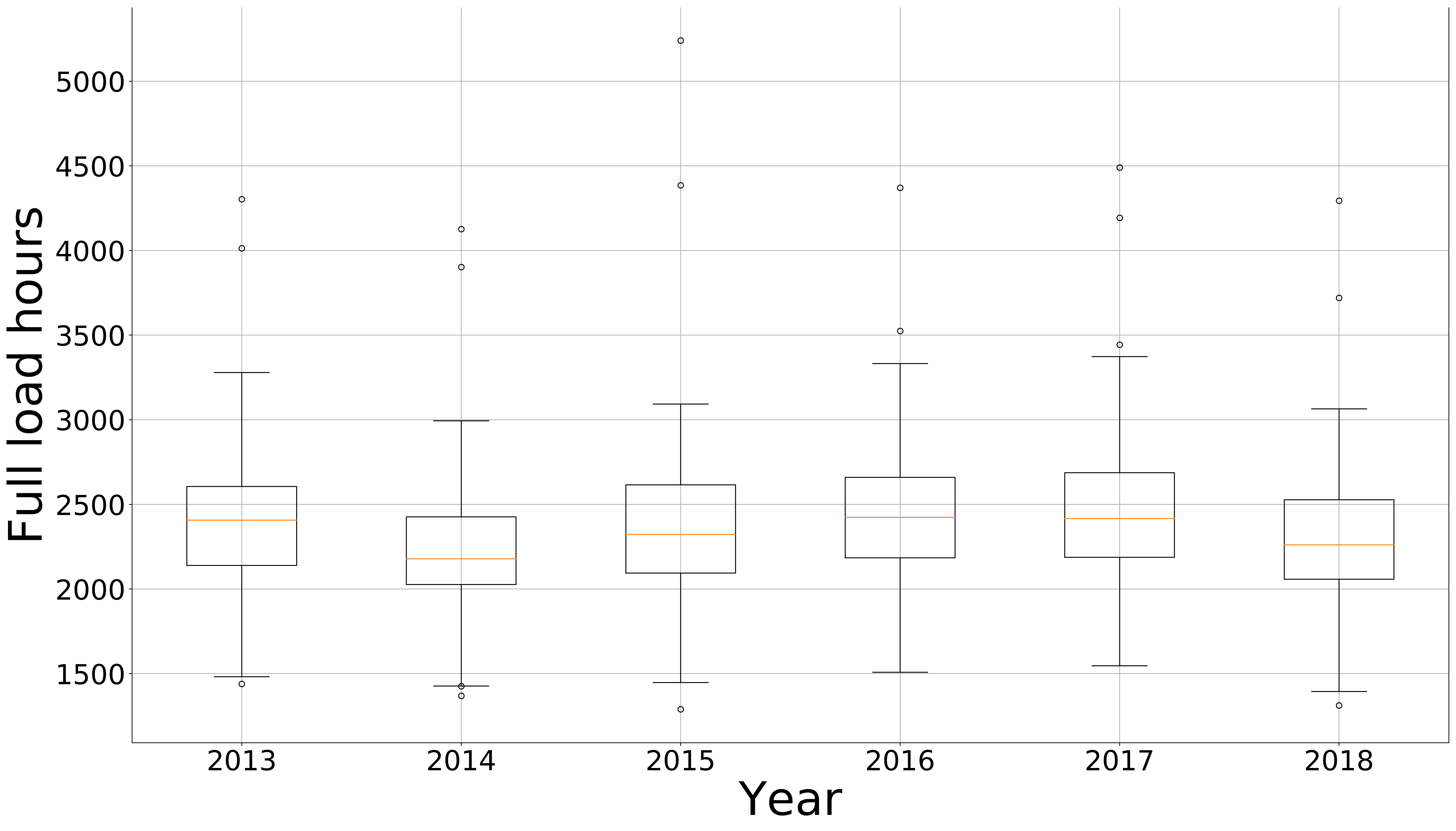}
    \caption{Boxplot of 84 customers' full load hours. The median is shown as the orange middle line. The box contains the 25 and 75\% quartiles, whereas the whiskers are 1.5 standard deviations. Outliers can be found outside the whiskers.}
    \label{fullloadhours}
\end{figure}

A box plot of the spread in full load hours of the consumers is shown in \Cref{fullloadhours}. Full load hours are defined as the total annual demand divided by the peak load. A high number indicates that the peak load is significantly higher than the average load and vice versa. Full load hours therefore indicate whether or not the customer has a flat, stable demand profile or few, large demand spikes. 

The median value is around 2300 hours for all years. It is lowest in 2014 and highest in 2017, the warmest and next coldest year, respectively, cf. \Cref{subsec:staticresults}. This looks counter intuitive as one would expect the highest demand in the coldest year. However, even warmer years have a few cold days resulting in high demand. On the other hand, cold years have high energy consumption, which reduces the full load hours, and consequently there is no direct relation between the lowest temperature and the number of full load hours.


\subsection{Grid tariff costs} \label{subsec:gridtariffcosts}

The underlying principle when setting the prices for the CS tariffs is that the income of the DSO remains the same after the transition from the present energy tariff. The annual fixed cost is set to the same for both CS tariffs. This is also the case for the energy term, which is set to the cost of marginal losses, estimated at about 0.5 €ct/kWh. We then vary the capacity cost until we find the level that results in approximately the same (aggregated) consumer costs as with the present tariff. All cost levels for CS and existing energy-based tariffs are shown in \Cref{SCcosts} and \Cref{ETcosts}, respectively. Finally, it is assumed that consumers do not adapt to the new tariff\footnote{This is, of course, a conservative assumption. By adapting behaviour, especially consumers with high peak demand, will save costs and become better off than shown in the subsequent analyses. The whole point of introducing capacity-based tariffs is to change consumer behaviour. Note that for dynamic capacity subscription, there is an imposed change in behaviour, i.e. keeping demand below the subscription limit during LLD activations.}.

The short-term consumer benefit does not change under the static CS tariff. It does change for the dynamic variant, which is taken into account through the calculation of the discomfort costs, cf. \Cref{subsec:discomfortcosts}. 

Correctly pricing the dynamic CS tariff is not straightforward, because the underlying principle of DSO cost recovery must be reconsidered. When demand is physically limited to the subscribed capacity, DSO revenues are reduced, because there is no payment for excess demand. On the other hand, over time, this has the potential of significantly reducing grid investments. We therefore argue that it is acceptable that DSO revenues decrease with dynamic capacity subscription, because costs will decrease over time. Instead, we consider that the consumers on aggregate should not be worse off when also taking into account the increase in consumer costs (or rather, reduction in benefit) caused by the demand reduction during LLD activation. The increase in consumer costs is calculated using the VCL model explained in \Cref{stocdynsubcapsection}. The consumer cost is also an expected value, as the number of activations is unknown in advance.

\begin{table}[h]
\centering
\caption{CS tariff costs}

\begin{tabular}{lcc}
\hline
\textbf{Cost element}       & \textbf{Cost} & \textbf{Unit} \\ \hline
\textbf{Fixed cost}         & 135           & $\euro/year$             \\
\textbf{Capacity cost (static)}      & 67.5          & $\euro/year$             \\
\textbf{Capacity cost (dynamic)}      & 54          & $\euro/year$             \\
\textbf{Energy term}        & 0.5           & $\euro ct/kWh$     \\
\textbf{Excess energy term} & 10            & $\euro ct/kWh$     \\
\textbf{Value of lost load}  & 5             & $\euro/kWh$       \\ \hline
\label{SCcosts}
\end{tabular}
\end{table}

\begin{table}[h]
\centering
\caption{Present energy tariff costs}

\begin{tabular}{lcc}
\hline
\textbf{Cost element} & \textbf{Cost} & \textbf{Unit} \\ \hline
\textbf{Fixed cost}   & 204.6         & $\frac{\euro}{year}$             \\
\textbf{Energy term}  & 1.859         & $\frac{\euro ct}{kWh}$    \\ \hline
\label{ETcosts}
\end{tabular}
\end{table}

\subsection{Activations}


As mentioned in \Cref{capacitytariffs}, we base activations on local congestions. We assume that the DSO activates the LLDs when the aggregated load of the 84 customers exceeds 390 kWh/h (peak load is 458 kWh/h). This is based on the idea that this would be the limit in the local grid, but the number is chosen to obtain results that illustrate the impact of the various tariff choices well. This means that we get a total of 291 hours of activation in 6 years, or 0.55\% of the time. We see from \Cref{activationtable} that more than half of the activations occur in 2015, whereas 2013 and 2016 have very few activations with only 13 and 10, respectively. 

\begin{table}[h]
\centering
\caption{Overview of activations for all years.}
\begin{tabular}{lccccccc}
\hline
\textbf{Activations} & \textbf{2013} & \textbf{2014} & \textbf{2015} & \textbf{2016} & \textbf{2017} & \textbf{2018} & \textbf{Total} \\ \hline
\textbf{Hours}       & 13 h          & 42 h          & 148 h         & 10 h          & 19 h          & 59 h          & 291 h          \\
\textbf{\% of hours} & 4.5\%         & 14.4\%        & 50.9 \%       & 3.4 \%        & 6.5 \%        & 20.3 \%       & 100\%          \\ \hline
\end{tabular}
\label{activationtable}
\end{table}



We used the same probability of occurring for each of the six years, because it is difficult to map those years convincingly on longer historical records of winter temperatures. Even if this could be accomplished, it it is not a good measure of the number of activations, as argued above. How probabilities should be allocated, to the number of activations is a key topic for further research, as this is an important parameter for the optimal subscription level.

%% file: Chapters/Results.tex
\section{Results \& discussion}\label{results}

An important question for capacity subscription is: to what level of capacity should a particular consumer subscribe? With perfect foresight on demand and activations, this is a simple optimisation problem, but in reality, demand and prices are, of course, unknown. Still, the perfect foresight solution can be used as a benchmark. We compare this with two other, realistic options:


\begin{itemize}
    \item Stochastic optimal subscribed capacity (stochastic optimum)
    \item Reactive subscribed capacity (reactive level).
\end{itemize}

The \textit{stochastic optimal subscribed capacity} is the subscription level resulting in the lowest expected annual cost under uncertainty. By considering uncertainty in domestic load, both high or low consumption profiles are taken into account when choosing a capacity level. 

The \textit{reactive subscribed capacity} is the subscribed capacity level based on the optimal level from the previous year. Essentially, this approach uses the exact same demand profile and activation pattern as from the previous year.


We find the stochastic optimum, based on the six scenarios with load profiles for 2013-2018. Temperatures dominate the load profiles in Norway due to high share of electric heating in households, making it important to have load data from both warm and cold winters to analyse the impact of the resulting consumption. 

As stated before, we only consider grid tariff costs, and ignore taxes, fees and electricity prices.

\subsection{Static capacity subscription} \label{subsec:staticresults}

\subsubsection{Stochastic approach} 

\begin{figure}[h]
    \centering
    \includegraphics[width=0.6\columnwidth]{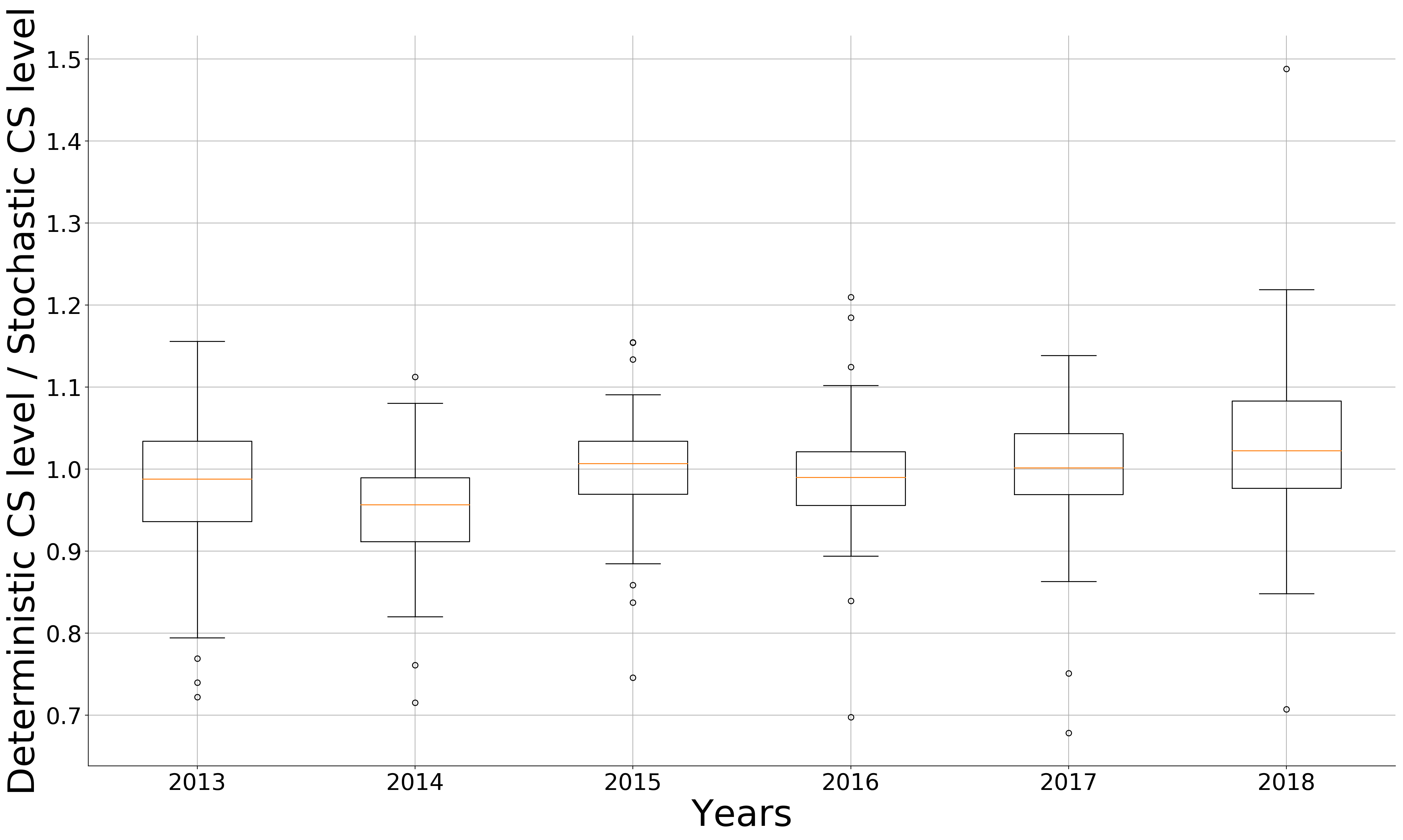}
    \caption{Boxplot of 84 customers' annual optimal subscribed capacity relative to the stochastic optimum.}
    \label{RelativeSubCap}
\end{figure}

Under static CS tariffs, the optimal subscribed capacity level results in the best trade-off between capacity costs and the excess energy term costs. Although there are some variations in the optimal subscribed capacity from year to year, the spread is not very large as seen in \Cref{RelativeSubCap}. The figure compares the annual deterministic optimum to the stochastic optimum. The mean values are in all cases within a 5\ margin, whereas the second and third quartile are in all cases within a 10\% margin of the stochastic optimum. The most extreme cases show deviations up to 50\%, but the whiskers (1.5 STD) mostly stay within a 2 \% margin.

\begin{figure}[h]
    \centering
    \includegraphics[width=0.6\columnwidth]{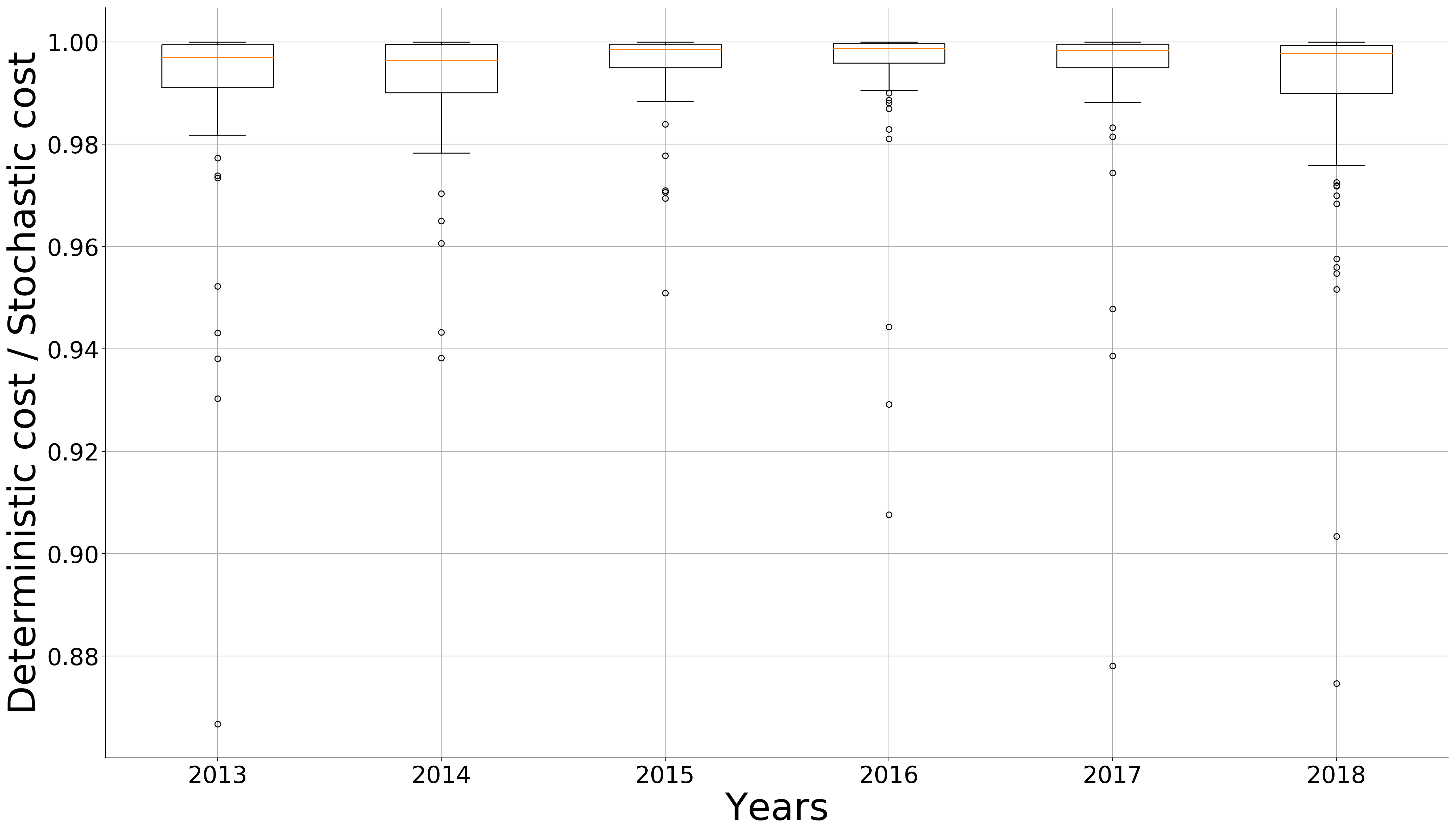}
    \caption{Box plot of 84 customers' annual cost when subscribing to the deterministic optimal level, relative to the stochastic optimal level.}
    \label{RelativeCost}
\end{figure}

Due to the perfect foresight, the deterministic optimum will result in lower costs than compared to stochastic optimum. The annual grid tariff costs under the deterministic optimum is therefore never higher than under the stochastic optimum, as shown in \Cref{RelativeCost}. However, the spread in costs is tiny. In almost all the cases, the cost when subscribing to the deterministic optimum and not the stochastic optimum is less than 3\% higher. Outliers show that the costs can deviate up to roughly 14\%, but this is rare. The spread in costs is surprisingly low compared to the spread in optimal subscribed capacity. However, this is fairly logical, as a higher subscription level results in high capacity costs and lower excess energy costs and vice versa, which is coherent with the results from \cite{Bjarghov2018}.


\begin{figure}[h]
    \centering
    \includegraphics[width=0.6\columnwidth]{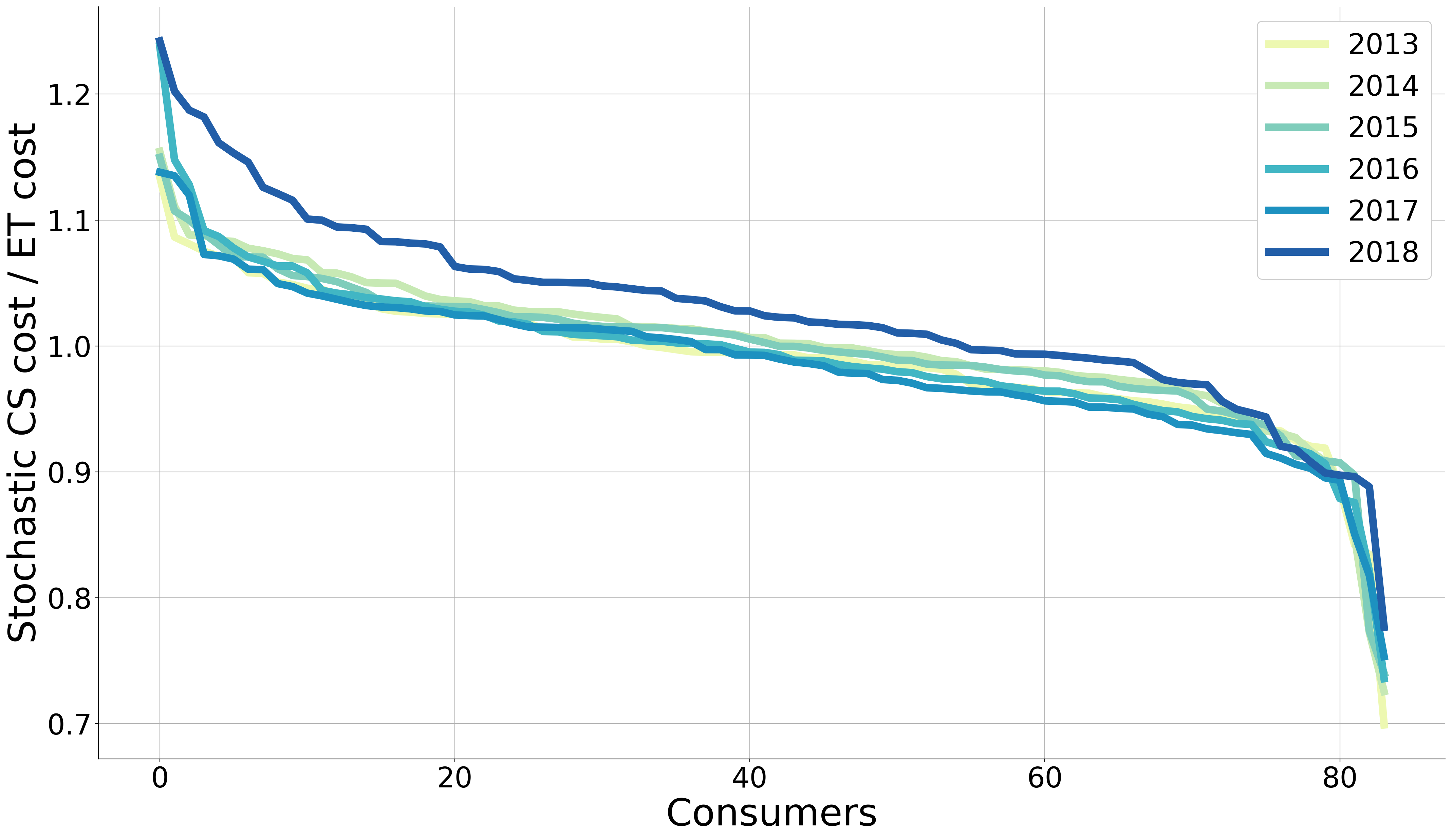}
    \caption{Sorted annual simulated stochastic customer cost relative to energy tariff costs.}
    \label{annualSimStocCost_vs_ETCost_duration}
\end{figure}

The sorted curve in \Cref{annualSimStocCost_vs_ETCost_duration} indicates how the new CS tariff compares to the existing energy tariff. The graph shows that the consumer annual costs increases the most in 2018. The 2016 costs reach similar maximum deviations, but not for as many consumers. These results imply that in 2018, the stochastic level is further away from the ex-post optimal deterministic level. This is confirmed in \Cref{RelativeSubCap}, which shows that a significant number of consumers would preferably subscribe to both lower and higher capacities (the spread is relatively large). However, the costs over the six years are the same (which is how the tariff cost level was set), meaning that costs simply deviate from year to year. This is further elaborated and discussed in \Cref{ReactivevsOpt}, which shows that costs differ from year to year, but not more than the existing energy tariff scheme.

\subsubsection{Reactive approach}

The stochastic approach requires several years of data in addition to being somewhat complicated. A more straightforward approach would be to use the data from the most recent year. As previously stated, we therefore use the term \textit{reactive subscribed capacity level}, which refers to using the optimal subscribed capacity of the previous year in the current year. For example, the reactive level corresponds to finding the optimal subscribed capacity of 2013 ex post and subscribing to that level in 2014.

\begin{figure}[h]
    \centering
    \includegraphics[width=0.6\columnwidth]{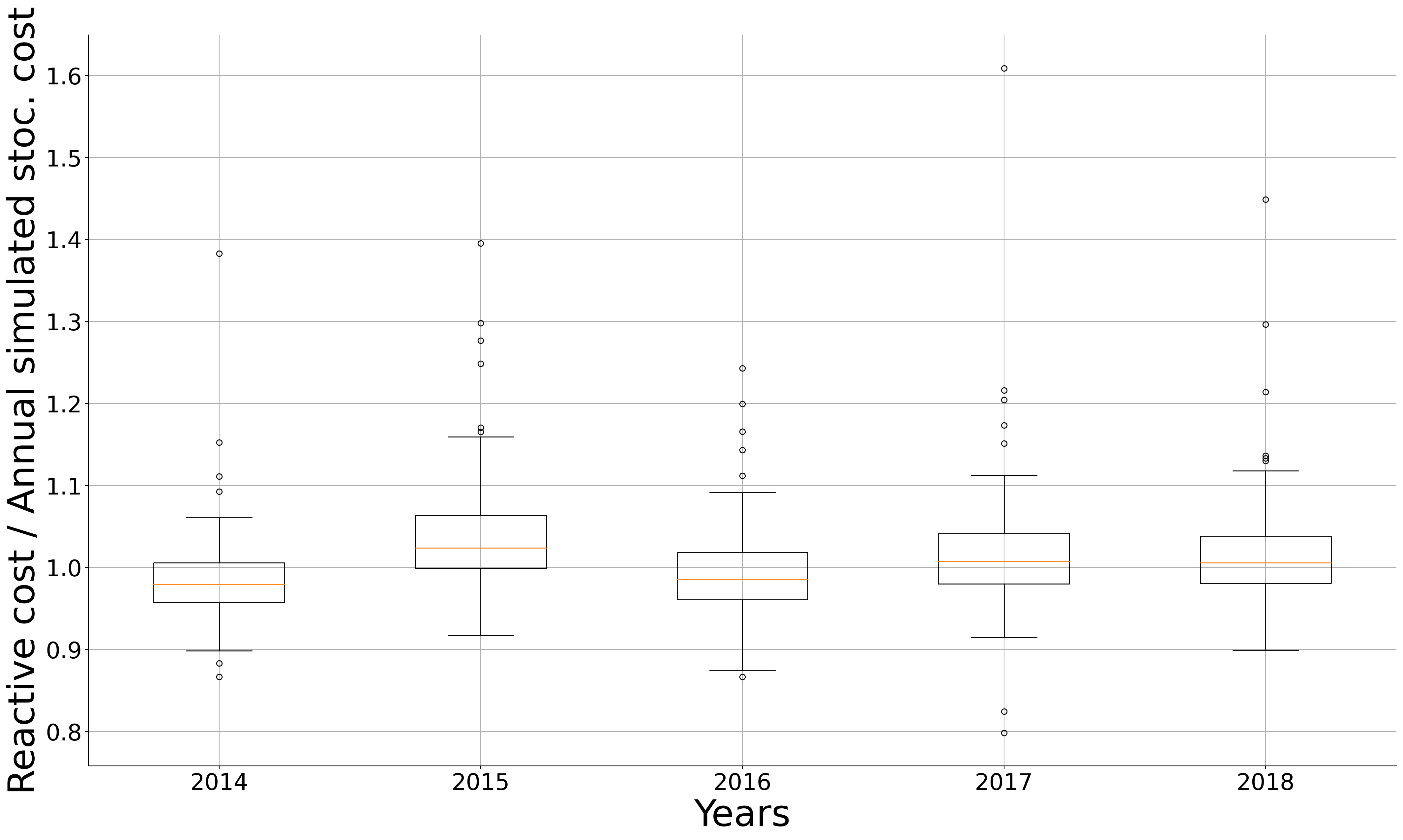}
    \caption{}
    \label{StochasticCost_vs_reactiveCost}
\end{figure}

The reactive level costs compared to the more robust stochastic approach mostly results in slightly higher costs on average. From \Cref{StochasticCost_vs_reactiveCost}, it can be deduced that subscribing to the wrong CS level mostly results in non-dramatic consequences as only outliers exceed an increased cost of 16\% compared to the stochastic approach. This is good news for DSOs that are afraid of their customers making sub-optimal choices instead of using the more robust stochastic approach. Outliers give up to 60\% increased costs.

\begin{figure}[h]
    \centering
    \includegraphics[width=0.6\columnwidth]{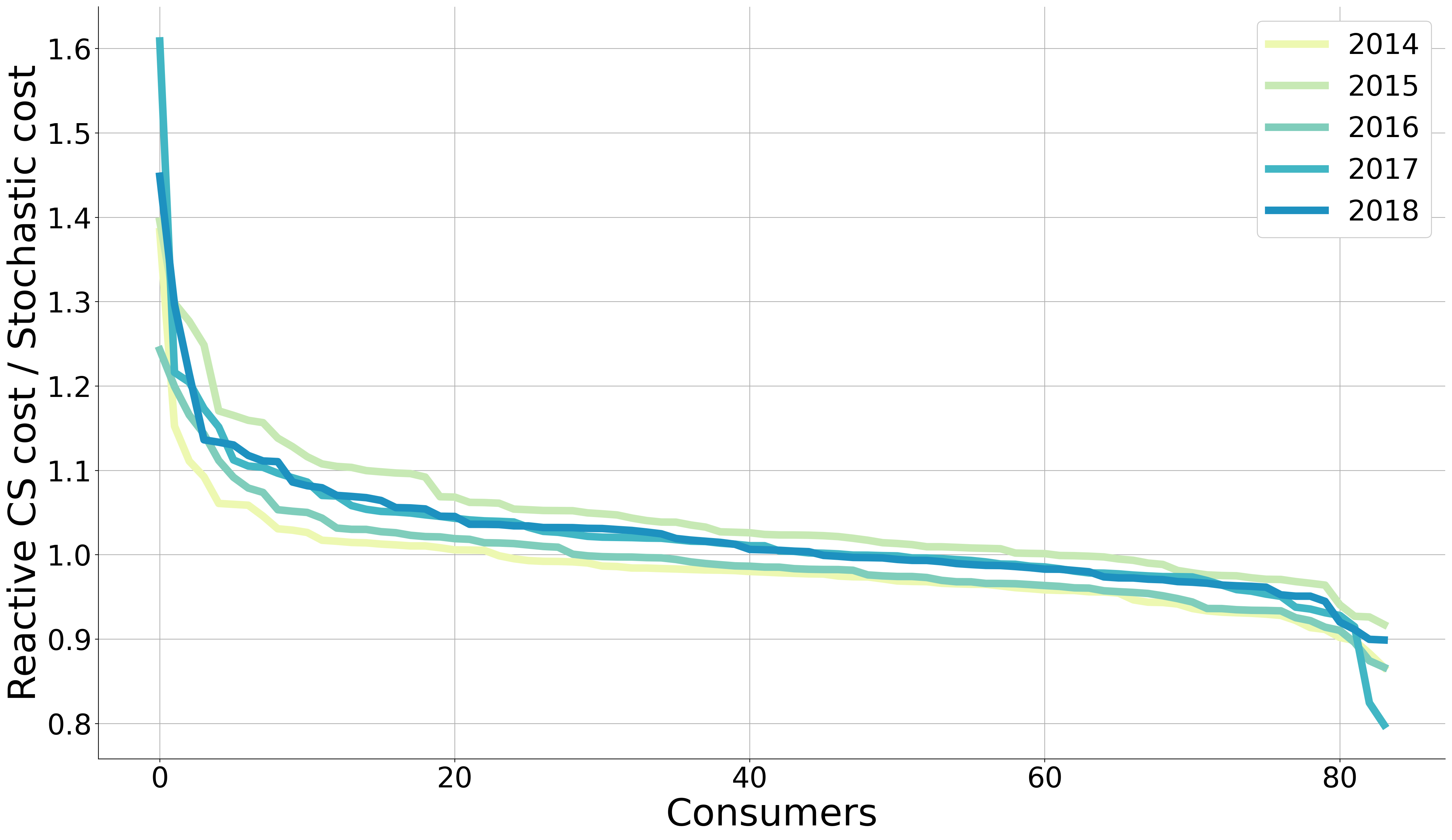}
    \caption{Sorted curve of reactive costs relative to annual simulated stochastic costs.}
    \label{annualSimStocCost_vs_ReactiveCost_duration}
\end{figure}

The same results as in \Cref{StochasticCost_vs_reactiveCost} is illustrated as a sorted curve in \Cref{annualSimStocCost_vs_ReactiveCost_duration}, where we see that acting reactively works relatively well for roughly 80-90\% of the consumers (who only experience up to +10\% cost increase), but results in high price increases for some consumers. 

\begin{figure}[h]
    \centering
    \includegraphics[width=0.6\columnwidth]{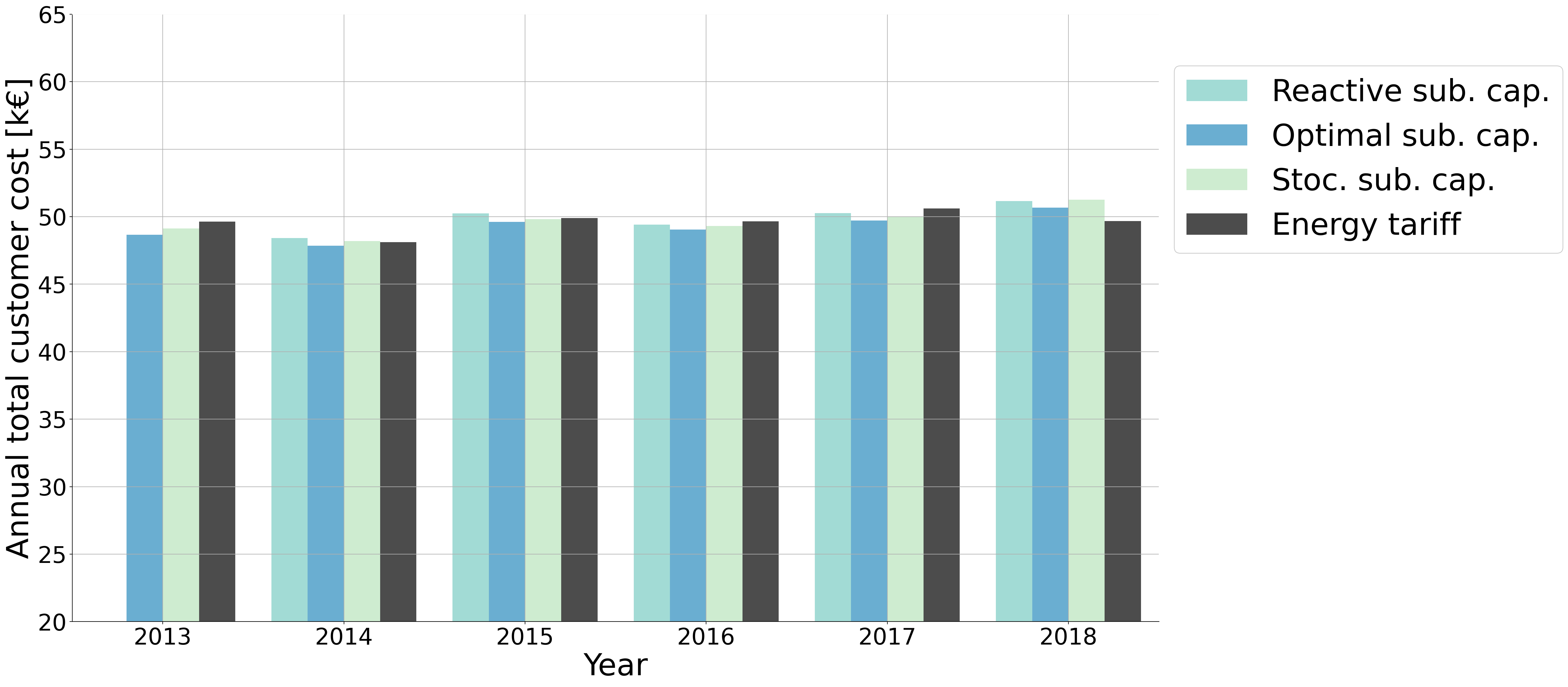}
    \caption{Aggregated annual consumer costs under the different approaches (static CS).}
    \label{ReactivevsOpt}
\end{figure}

Looking at the results from a distance, \Cref{ReactivevsOpt} shows that costs increase by 1.2 - 2.0 \% for the total customer group when always subscribing to the reactive subscribed capacity level compared to the deterministic optimum. The figure also shows that there is a significant variation in annual costs, regardless of the tariff. This is good for the DSO, who is interested in predictable cost-recovery, but also for customers who should not experience increasing cost fluctuations with CS tariffs. Of the CS tariffs, the deterministic optimum is obviously always lowest, whereas the stochastic optimum mostly gives lower costs than the reactive, except for 2018 when the reactive subscribed capacity level gives slightly better results. This exception occurs if the demand profiles from two years match relatively well, and both deviate somewhat from the stochastic level. In general, the spread from year to year is relatively small and does not vary more than the existing energy tariff.


\subsection{Dynamic capacity subscription}

Under the dynamic CS tariff scheme, the load profile characteristics in terms of seasonal variations, flatness and ``spikiness'' significantly influence the results because activation of the LLD only causes discomfort costs to the consumer if the system peak correlates with the consumer peak. The DSO forecasts demand peaks and sends activations based on expected grid congestions. As peak load hours occur when demand is high, it is natural to assume that the correlation between consumer peak loads and neighbourhood peak loads are high. However, this is not always the case for the individual customer. Customers with flatter load profiles and/or customers with other heating sources than electricity do not necessarily share peak loads with the system. Customers with low correlation between individual peaks and activations will therefore be able to subscribe to relatively low capacities because the discomfort costs during activations are low. Flexible customers who can reduce load during activations will also be rewarded with the dynamic CS tariff. We do not model any flexibility assets in this study, but model demand flexibility implicitly by curtailing some load under the dynamic CS tariff scheme.

\subsubsection{Stochastic approach}

\begin{figure}[h]
    \centering
    \includegraphics[width=0.6\columnwidth]{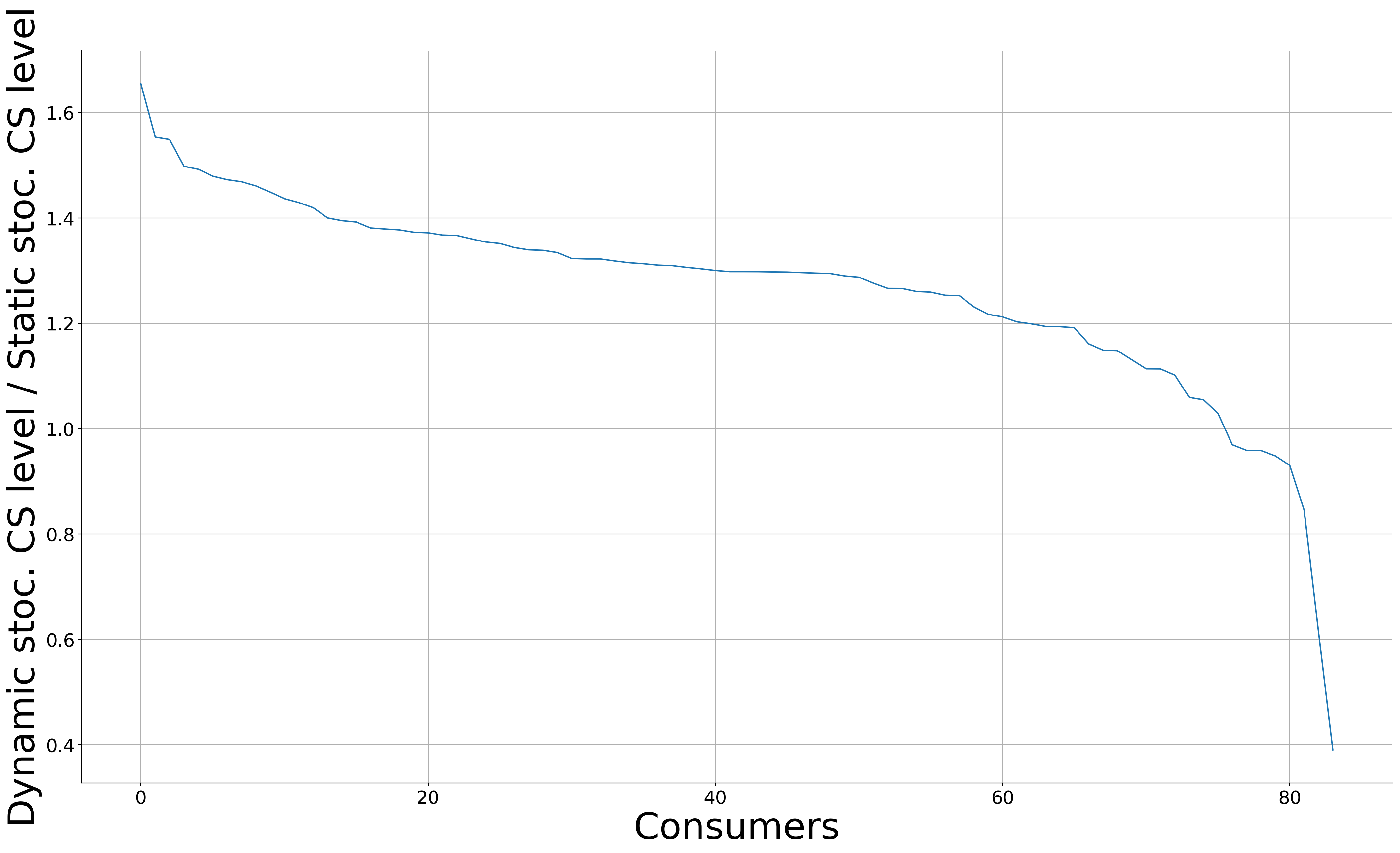}
    \caption{Stochastic dynamic CS level relative to stochastic static CS level.}
    \label{DynamicStocLevel_vs_StaticStocLevel_duration}
\end{figure}

Results show that customers subscribe to significantly higher capacities under the dynamic compared to the static CS tariff, with the median increase roughly 30\% higher, as seen in \Cref{DynamicStocLevel_vs_StaticStocLevel_duration}. This is mostly due to the difference in excess energy cost, which is 0.1 \EUR{} in the static case, but up to 5 \EUR{} (VoLL) in the dynamic case because the customers are physically limited, cf. \Cref{VCLpercent}. The capacity cost is also somewhat lower in the dynamic case (adjusted to match the DSO cost recovery). The spread is relatively large, with some customers preferring to subscribe to as little as 40\% under the dynamic compared to the static CS tariff, indicating that their peak loads are not coinciding with activations, or that their load profiles are flat. This stands in contrast to some exceptions on the other side of the scale, where two customers subscribe to more than 60\% more in the dynamic case, indicating a high correlation between activations and peak loads. It is therefore natural that they would subscribe to more (and thus pay more) because activations result in more load shedding. Due to the heavy variation in deterministic dynamic CS levels, the importance of using a stochastic approach is clear.

\begin{figure}[h]
    \centering
    \includegraphics[width=0.6\columnwidth]{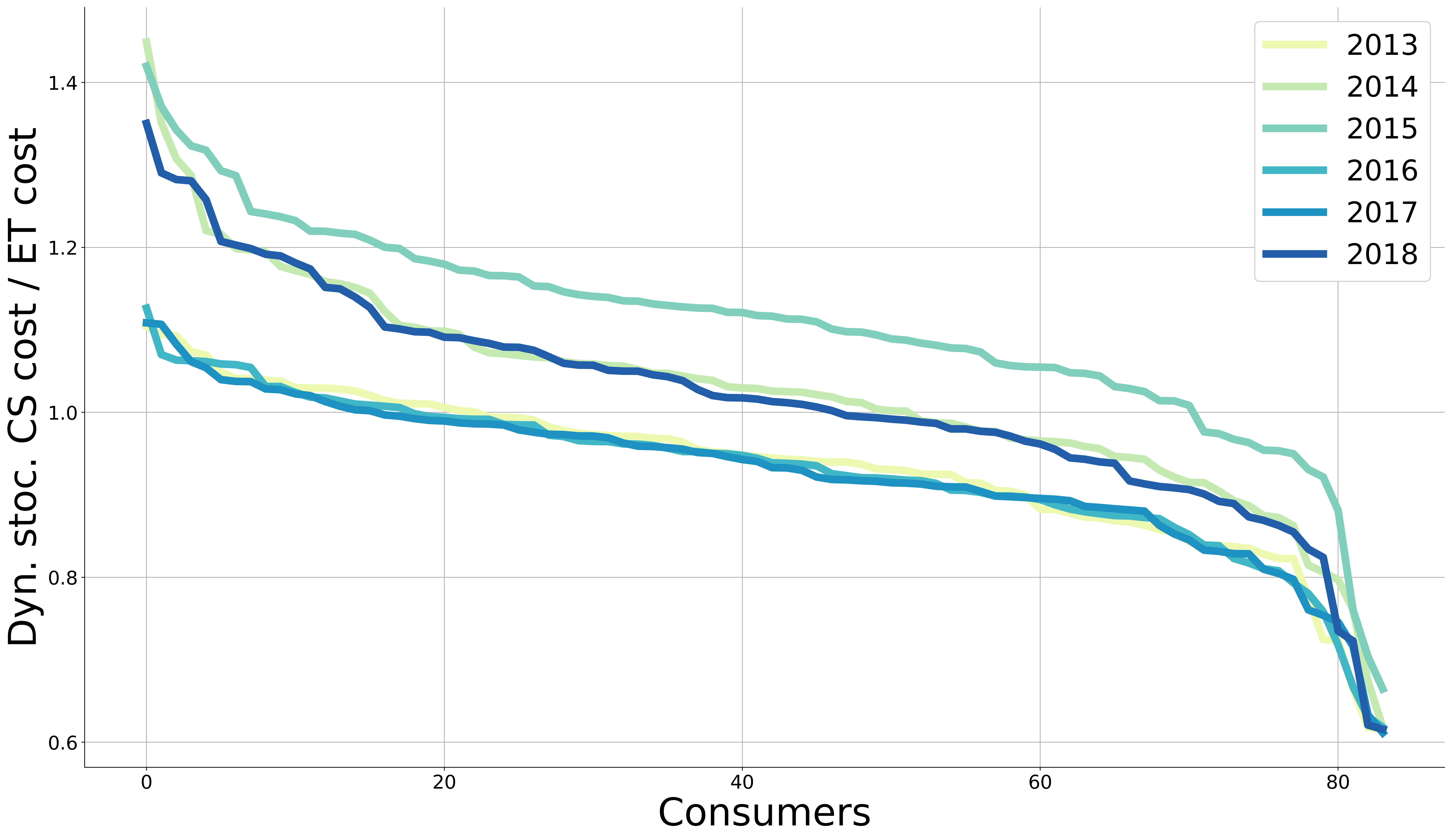}
    \caption{Annual simulated dynamic costs compared to energy tariff costs.}
    \label{annualSimDynStocCost_vs_ETCost_duration}
\end{figure}

Compared to the old energy tariff, \Cref{annualSimDynStocCost_vs_ETCost_duration} shows a significant spread in annual costs under the dynamic CS tariff. 2015 especially has higher costs due to the high number of activations leading to higher VCL. 


\subsubsection{Reactive approach}

In general, the deterministic optimal subscribed capacity level under the dynamic CS level varies much more from year to year due to the difference in number of activations. In a year with few activations, the optimal subscription level can be as low as zero or close to zero because of the low VCL cost. On the other hand, years with many activations results, in a high optimal deterministic subscription level. This shows that the reactive approach cannot be used for dynamic CS.

\begin{figure}[h]
    \centering
    \includegraphics[width=0.75\columnwidth]{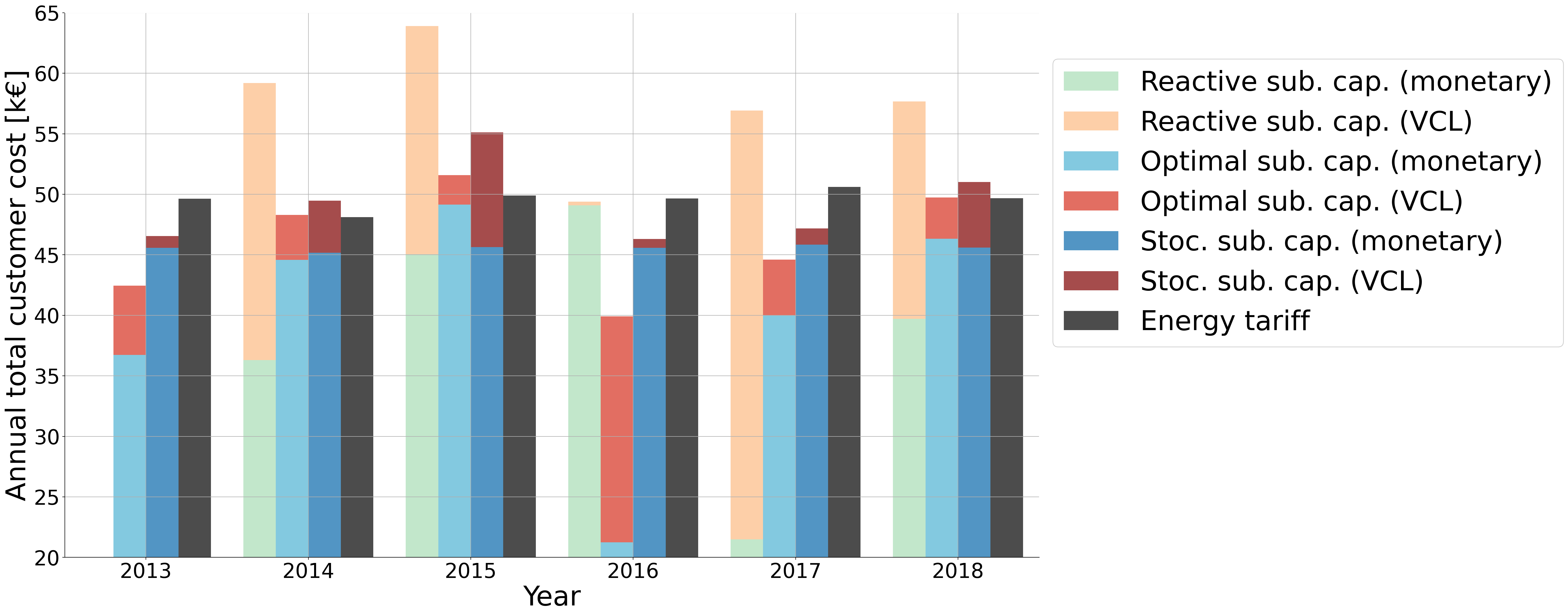}
    \caption{Aggregated annual consumer costs under the different approaches (dynamic CS).}
    \label{DynReactivevsOpt}
\end{figure}

In \Cref{DynReactivevsOpt}, the aggregated annual consumer costs are shown. The blue bars (bottom) represent actual monetary costs (which corresponds to the DSO's income), whereas the red bars on top of the blue are discomfort costs related to the value of cut load and are not monetary costs. The costs under the existing energy tariff scheme are shown as black bars. Acting reactively clearly results in the highest costs as consumers subscribe to sub-optimal levels. This leads to either extremely high discomfort costs if they have insufficient subscribed capacity (as shown in 2017), or close to zero discomfort costs due to subscribing to a high capacity (as shown in 2016). Both cases result from reacting to few or many load limitation activations from the previous year. This also leads to unacceptable variations in the DSO's revenues. When using the stochastic approach, the costs are more stable and relatively similar to the energy tariff costs. The average total cost (monetary + discomfort) are the same (by calibration), but this results in somewhat lower revenues to the DSO. In the long run, this seems acceptable, as dynamic CS probably is very efficient in reducing peak demand, reducing the need for grid investments. Under the stochastic approach, the DSO income is also relatively stable and does not vary more than the energy-based tariff scheme. This is good news for DSOs who rely on stable income. The theoretical optimal costs are of course lower than the other approaches.

\begin{figure}[h]
    \centering
    \includegraphics[width=0.6\columnwidth]{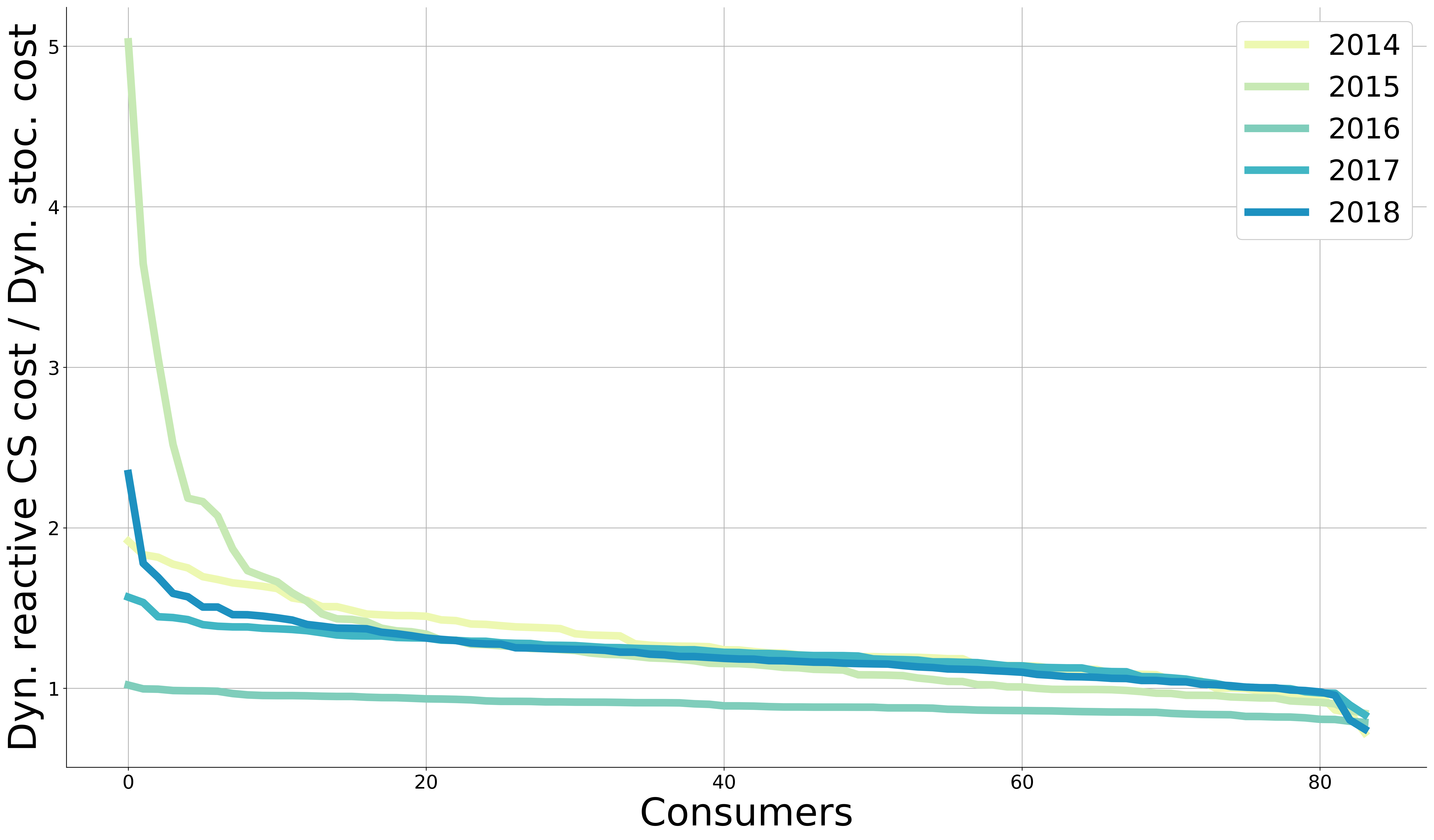}
    \caption{Dynamic CS reactive costs relative to annual simulated dynamic CS costs.}
    \label{annualSimDynStocCost_vs_DynReactiveCost_duration}
\end{figure}

\Cref{annualSimDynStocCost_vs_DynReactiveCost_duration} in particular clearly illustrates that reactive determination of the subscription level is a strategy that cannot be used for dynamic capacity subscription. More advanced strategy like the proposed stochastic strategy are necessary and are no impediment to implementation of dynamic capacity subscription given the present availability of data and support tools on smart phones.


\subsection{Consequences for consumers and the DSO}

\begin{figure*}[t!]
    \centering
    \begin{subfigure}[b]{0.45\textwidth}
        \centering
         \includegraphics[width=\textwidth]{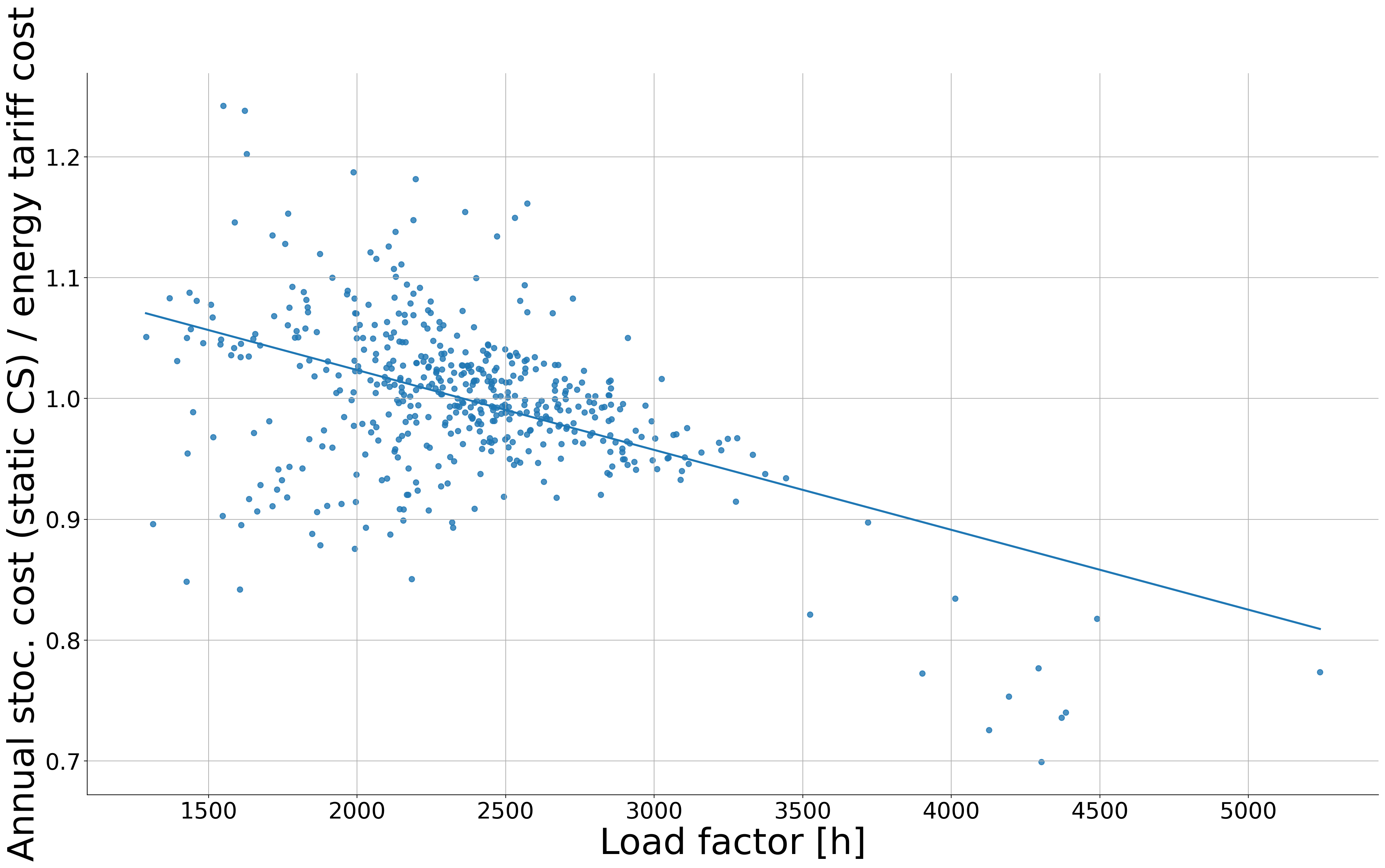}
         \caption{Static CS tariff.}
         \label{fig:Correlation_stochasticcost_vs_utime}
    \end{subfigure}%
    \begin{subfigure}[b]{0.45\textwidth}
        \centering
         \includegraphics[width=\textwidth]{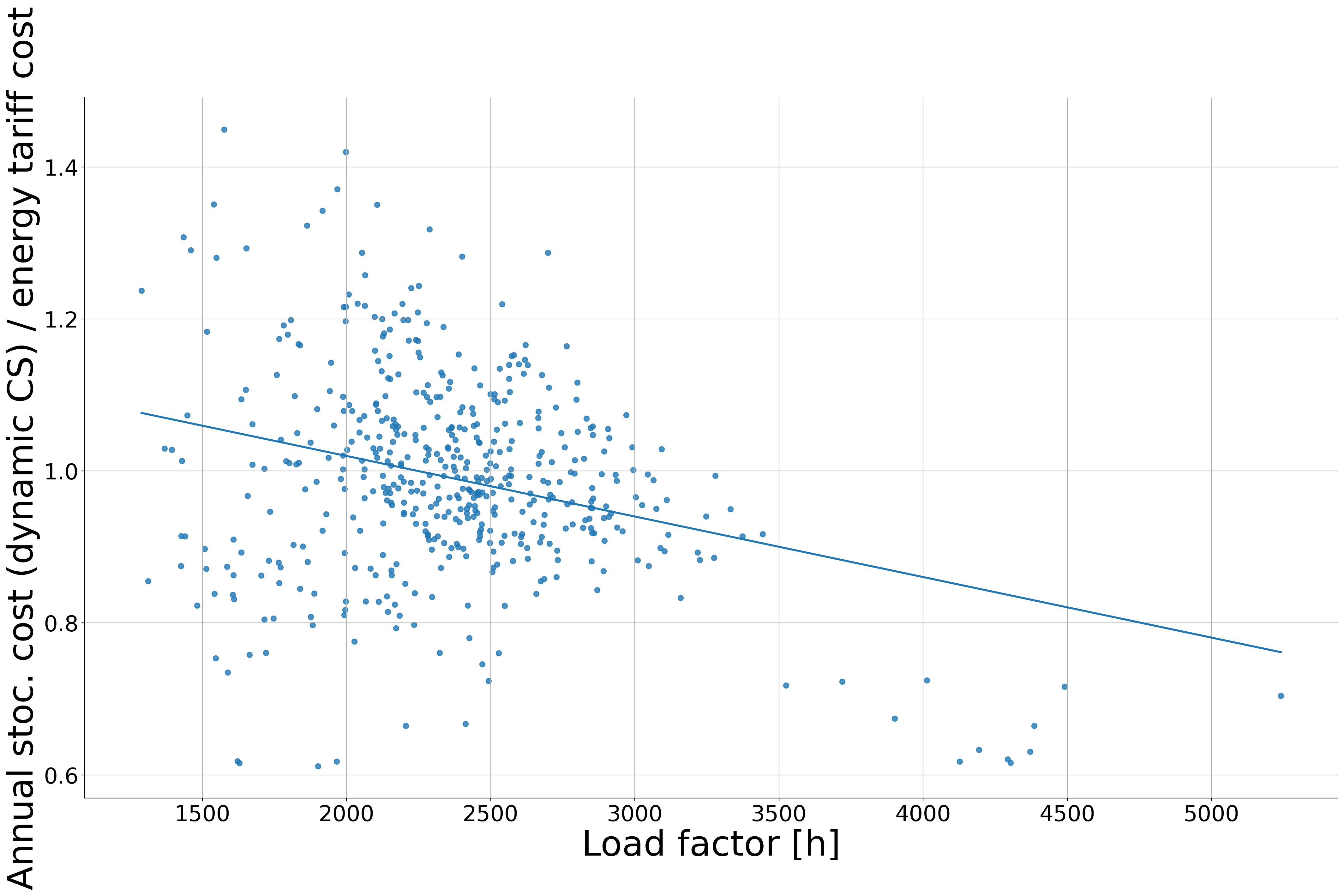}
         \caption{Dynamic CS tariff.}
         \label{fig:Correlation_dynstochasticcost_vs_utime}
    \end{subfigure}
    \caption{Scatter plots of static and dynamic CS costs compared to the energy-based tariff, in relation to the load factor. Linear regression is shown as blue lines.}
\end{figure*}

The correlation between load factor and cost-redistribution under the static and dynamic CS tariffs is shown in \Cref{fig:Correlation_stochasticcost_vs_utime} and \Cref{fig:Correlation_dynstochasticcost_vs_utime}, respectively. The load factor is similar to full load hour term used for power production, and represents how steep the peak load is compared to the annual consumption. A low and high load factor implies that the peak load is high or low compared to the annual consumption, respectively. A high load factor is associated with flat low profiles which should be rewarded by CS tariffs. This trend is shown clearly in the results. However, there is a spread in the data, which stems from the fact that peak load is not always the deciding factor. If a consumer has a high peak load in just a few hours, but a flat profile otherwise, this is not penalised as heavily by the CS tariffs. In the dynamic CS case, the spread is even larger as it also consider system peaks. Consumers with peak loads outside of the system peak loads are not penalised as heavily as those who coincide with system peaks. 

Some of the negative impacts of consumer versus system peak coincidence could be improved by only activating load limitation in parts of the grid where there is scarcity. However, this is not allowed in Norway due to fairness principles. This challenging compromise between cost-reflectivity and fairness could be solved by the regulators. To achieve this, the regulator and DSOs could investigate methods to properly compensate consumers that are located in areas with more frequent load limitations, for example in the form of reduces fixed costs. This is of increasing importance in countries where there are few, large DSOs, with many customers. In those countries, it would be very inefficient to limit load on all of them.

Currently, consumers with high peak loads are being subsidised by consumers with flatter load profiles that have lower contributions to system peaks. A potential move from energy-based to capacity-based network tariffs has significant cost-reflectiveness advantages, as consumers would have an incentive to reduce their peak load. As shown in \Cref{DynReactivevsOpt}, the DSO income is somewhat reduced when moving to dynamic CS tariffs as the cost is shifted onto the consumer in the shape of discomfort costs\footnote{The DSO income is the same when discomfort costs are included, but because they are not monetary costs, the income is somewhat reduced.}. This could be difficult for DSOs in the short-term who rely on steady income. In the case of a move towards a dynamic CS tariff, we therefore recommend a somewhat slower transition in terms of reducing the grid tariff prices until the DSOs starts to see lower grid investment costs. This transition period also benefits from the fact that the more data, the better advice to consumers can be given. In general, \Cref{ReactivevsOpt} and \Cref{DynReactivevsOpt} shows that DSOs can not expect the variance in annual income to increase(given that the stochastic approach is applied), and should thus not be used as an argument against moving towards capacity-based tariffs.  

%% file: Chapters/Conclusion.tex
\section{Conclusions \& policy implications} \label{conclusion}

This paper demonstrates the change in annual grid tariff costs for a sample of household customers when applying capacity-based tariff structures. Two types of capacity subscription tariffs were analysed; static and dynamic subscribed capacity. Under the static capacity subscription tariff, results show that using a stochastic approach to determine the subscribed capacity level using several years with historical consumption data results in annual costs close to the perfect foresight theoretical optimum. Acting reactively (based on the previous year's conditions), works reasonably well for most consumers, but leads to significantly increased costs for a few consumers. The DSO cost recovery is equally stable as under the energy tariff, with tiny variations from year to year. Under the dynamic CS tariff, consumers are only limited during hours with grid scarcity. The stochastic approach is significantly better than acting reactively as the number of activations from year to year is very different. Subscribing reactively leads to huge variations in subscribed capacity from year to year, resulting in unacceptable demand limitations and wide variations in annual DSO revenues. This approach cannot be allowed used in practice. This can e.g. to some extent be avoided by requiring a minimum subscription level.

Overall, the static CS tariff results in low to moderate changes in annual costs for consumers, is robust to sub-optimal subscription levels and does not result in increased variance in costs compared to the existing energy tariff. Regulators should consider moving to such tariffs in the future, as capacity subscription tariffs benefit from being more cost-reflective while maintaining a stable DSO income. Advising consumers on optimal subscription levels is also fairly easy with the suggested method, implying that regulator/DSO should be able to help consumers find a reasonable subscription level. The tariffs also redistribute costs between consumers based on their peak loads, removing some cross-subsidisation from consumers with low peak loads to consumers with high peak loads.


The impact of the dynamic CS tariff are more difficult to assess, because in addition to the actual payments from consumer to DSO, also the loss of consumer welfare due to demand limitations need to be taken into account. We use a simple, assumed cost function for this effect. On the other hand, the dynamic CS tariff offers a much more precise limitation on load which is more efficient, as no load limitations or excess energy fees exist during hours with no grid scarcity. The monetary costs for the consumers are relatively stable but somewhat lower. In the case of a transition to dynamic capacity-subscription tariffs, the regulator should consider a transition period before the grid tariff prices are reduced according to the reduced future grid investments. In order to avoid load limitation on many consumers in large DSO areas, the regulator should look into methods to compensate consumers who are frequently limited, which would increase overall efficiency of the tariff.

Further work might look deeper in the effect on particular consumer segments, based on customer type data and heating sources, which could have shed extra light on what type of consumers experience different cost impacts. Future work is recommended to look more extensively into the impact of activation scenarios, adding consumer flexibility and addressing the consumer types implications. A method to estimate optimal subscription for dynamic capacity subscription if no previous data (or only short period) is available should also be of interest for further work. Moreover, the dynamic CS tariff could be extended to only limiting load in grid areas with congestions. However, this may not be allowed under existing regulation.

%% file: Chapters/Acknowledgements.tex
\section*{Acknowledgement}

This work was funded by the ``DigEco - Digital Economy'' project funded by the NTNU Digital Transformation Initiative (project number: 2495996). We would like to thank Roman Hennig and Laurens de Vries at TU Delft, and Matthias Hofmann at Statnett for their useful input and contributions.

%% file: Chapters/Appendix.tex
\appendix
\appendixpage
\addappheadtotoc

\begin{appendices} \label{Appendix}

\end{appendices}